\documentclass[iop, twocolumn]{aastex63}
\hypersetup{citecolor=blue,urlcolor=red}
\shorttitle{Black widow PSR J1720-0533}
\shortauthors{Wang et al.}

\begin{document}

\title{Unusual Emission Variations Near the Eclipse of A Black Widow PSR J1720$-$0533}

\correspondingauthor{J.B. Wang, N. Wang}
\email{wangjingbo@xao.ac.cn, na.wang@xao.ac.cn}

\author{S.Q. Wang}
\affiliation{Xinjiang Astronomical Observatory, Chinese Academy of Sciences, Urumqi, Xinjiang 830011, People's Republic of China}
\affiliation{CAS Key Laboratory of FAST, NAOC, Chinese Academy of Sciences, Beijing 100101, People's Republic of China}

\author{J.B. Wang}
\affiliation{Xinjiang Astronomical Observatory, Chinese Academy of Sciences, Urumqi, Xinjiang 830011, People's Republic of China}
\affiliation{Key Laboratory of Radio Astronomy, Chinese Academy of Sciences, Urumqi, Xinjiang, 830011, People's Republic of China}
\affiliation{Xinjiang Key Laboratory of Radio Astrophysics, Urumqi, Xinjiang, 830011, People's Republic of China}

\author{N. Wang}
\affiliation{Xinjiang Astronomical Observatory, Chinese Academy of Sciences, Urumqi, Xinjiang 830011, People's Republic of China}
\affiliation{Key Laboratory of Radio Astronomy, Chinese Academy of Sciences, Urumqi, Xinjiang, 830011, People's Republic of China}
\affiliation{Xinjiang Key Laboratory of Radio Astrophysics, Urumqi, Xinjiang, 830011, People's Republic of China}

\author{J.M. Yao}
\affiliation{Xinjiang Astronomical Observatory, Chinese Academy of Sciences, Urumqi, Xinjiang 830011, People's Republic of China}
\affiliation{Key Laboratory of Radio Astronomy, Chinese Academy of Sciences, Urumqi, Xinjiang, 830011, People's Republic of China}
\affiliation{Xinjiang Key Laboratory of Radio Astrophysics, Urumqi, Xinjiang, 830011, People's Republic of China}

\author{G. Hobbs}
\affiliation{CSIRO Astronomy and Space Science, PO Box 76, Epping, NSW 1710, Australia}

\author{S. Dai}
\affiliation{School of Science, Western Sydney University, Locked Bag 1797, Penrith South DC, NSW 2751, Australia}

\author{F.F. Kou}
\affiliation{Xinjiang Astronomical Observatory, Chinese Academy of Sciences, Urumqi, Xinjiang 830011, People's Republic of China}
\affiliation{Key Laboratory of Radio Astronomy, Chinese Academy of Sciences, Urumqi, Xinjiang, 830011, People's Republic of China}
\affiliation{Xinjiang Key Laboratory of Radio Astrophysics, Urumqi, Xinjiang, 830011, People's Republic of China}

\author{C.C. Miao}
\affiliation{National Astronomical Observatories, Chinese Academy of Sciences, Beijing 100101, People's Republic of China}

\author{D. Li}
\affiliation{National Astronomical Observatories, Chinese Academy of Sciences, Beijing 100101, People's Republic of China}
\affiliation{NAOC-UKZN Computational Astrophysics Centre, University of KwaZulu-Natal, Durban 4000, South Africa}

\author{Y. Feng}
\affiliation{Zhejiang Lab, Hangzhou, Zhejiang 311121, China}

\author{S.J. Dang}
\affiliation{School of Physics and Electronic Science, Guizhou Normal University, Guiyang, 550001, People's Republic of China}

\author{D.H. Wang}
\affiliation{School of Physics and Electronic Science, Guizhou Normal University, Guiyang, 550001, People's Republic of China}

\author{P. Wang}
\affiliation{National Astronomical Observatories, Chinese Academy of Sciences, Beijing 100101, People's Republic of China}

\author{J.P. Yuan}
\affiliation{Xinjiang Astronomical Observatory, Chinese Academy of Sciences, Urumqi, Xinjiang 830011, People's Republic of China}
\affiliation{Key Laboratory of Radio Astronomy, Chinese Academy of Sciences, Urumqi, Xinjiang, 830011, People's Republic of China}
\affiliation{Xinjiang Key Laboratory of Radio Astrophysics, Urumqi, Xinjiang, 830011, People's Republic of China}

\author{C.M. Zhang}
\affiliation{National Astronomical Observatories, Chinese Academy of Sciences, Beijing 100101, People's Republic of China}

\author{L. Zhang}
\affiliation{National Astronomical Observatories, Chinese Academy of Sciences, Beijing 100101, People's Republic of China}

\author{S.B. Zhang}
\affiliation{Purple Mountain Observatory, Chinese Academy of Sciences, Nanjing 210008, People's Republic of China}

\author{W.W. Zhu}
\affiliation{National Astronomical Observatories, Chinese Academy of Sciences, Beijing 100101, People's Republic of China}

\begin{abstract}

We report on an {unusually} bright observation of PSR J1720$-$0533 using the Five-hundred-meter Aperture Spherical radio Telescope (FAST). The pulsar is in a black widow system that {was discovered by the Commensal Radio Astronomy FAST Survey (CRAFTS). 
By coincidence, a bright scintillation maximum was simultaneous with the eclipse in our observation which allowed for precise measurements of flux density variations, as well as dispersion measure (DM) and polarization.} 
We found that there are quasi-periodic pulse emission variations with a modulation period of $\sim$ {22\,s} during the ingress of the eclipse, which could be caused by plasma lensing. {No such periodic modulation was found during the egress of the eclipse. }
{The linear polarization of the pulsar disappears before the eclipse, even before there is a visually obvious change in DM. We also found that the pulse scattering maybe play an important role in the eclipse of PSR J1720$-$0533.}

\end{abstract}


\section{INDRUCTION}

Redbacks (RB) and  black widows (BW), jointly known as the spider pulsars,  are an interesting subset of the pulsar population~\citep{Roberts13, Patruno17}.
Spider pulsar systems comprise a millisecond pulsar (MSP) with a low-mass companion in short, near-circular orbits. The two types are distinguished by the masses of their companions, { with RB companion masses $\sim$0.2$-$0.4 $M_{\odot}$, and BW companion masses $\sim$0.01$-$0.05 $M_{\odot}$~\citep{Roberts13}}. Most of known spider systems show periodic eclipses of the radio emission centred approximately around inferior conjunction of the companions. The eclipses happen when the radio emission is blocked by the companions or {outflowing material}. 
The presence of material beyond the Roche lobes and strong irradiation indicate the companion will be ablated by the emission from the neutron star~\citep{Fruchter88}.

The eclipses of spider pulsars are frequency dependent, usually  with longer duration at lower frequency and sometimes no eclipses are seen at all at high-frequency~\citep{Stappers01,Polzin18, Polzin19}. Although many eclipse mechanisms have been proposed, there is no apparent consensus on the correct eclipse mechanism. Different mechanisms may be responsible for the eclipses in different systems~\citep{Thompson94}.  
Magnetic fields in the eclipse medium are required for some of the promising mechanisms. {The presence of a magnetic field} of the eclipse material is suggested by the linear depolarization near eclipse for a RB PSR J1748$-$2446A~\citep{You18}. \citet{Crowter20} measured a non-zero magnetic field in the eclipse material of a BW PSR J2256$-$1024. 

Spider pulsars are considered to be descendants of low-mass X-ray binaries (LMXBs) after accretion on to the pulsar has terminated~\citep{Bhattacharya91}. Systems with transitions between LMXBs and radio pulsars have been seen in three RBs (e.g. \citealt{Archibald09, Stappers14}), {which have strongly inspired} the typical MSP recycling theory~\citep{Alpar82}. 
The ablation of a companion after accretion may lead to complete destruction of the companion star, contributing to the observed isolated MSPs~\citep{Fruchter88}. 
Apart from evolutionary studies, spider pulsars could also offer valuable opportunities to investigate the pulsar wind and characteristics of the companion stars under intense irradiation.

Plasma lensing was detected surrounding eclipses in three spider pulsars in the past few years, BW PSR B1957+20 ~\citep{Main18}, RB PSR B1744$-$24A~\citep{Bilous19}, and BW PSR 2051$-$0827~\citep{Lin21}. The effects of plasma lensing were seen as highly magnified pulses. 
The plasma lensing was used to resolve the pulse emission, constraining emission sizes and separations. 
\citet{Main18} inferred a resolution of the plasma lensing of PSR B1957+20 of about 10\,km, which is comparable to the pulsar’s radius.
{ However, this method is limited by understanding and modelling of the lenses. }

The radio emissions of the spider pulsars are generally weak near the eclipses. {Highly sensitive} radio telescopes such as the Five hundred metre Aperture Spherical Telescope (FAST) provides a great opportunity to study the eclipse in detail.
PSR J1720$-$0533 is a pulsar in the Galactic field {which was} newly discovered by the Commensal Radio Astronomy FAST Survey (CRAFTS, see~\citealt{Li18}) and was confirmed to be a BW by FAST key science project: pulsar physics and evolution (project id: ZD2020\_6) with a 3.26\,ms spin period, a 3.16\,hr orbital period and {a $\sim$0.034\,$M_{\odot}$ companion}.
The dispersion measure (DM) is $36.8337\pm0.0006\,\rm cm^{-3}\,pc$ with the DM-derived distance of $191$\,pc using the electron density model of~\citet{Yao17}. 
The timing solution for the pulsar will be presented by {\color{blue} Miao et al, 2021}.
In the paper, we present the unusual emission variations near the eclipse of PSR J1720$-$0533 using FAST.
In Section 2, we {describe } our observation  and data processing. In Section 3, we {present} the results.  We {discuss} and {summarize} our results in Section 4.

\section{OBSERVATIONS AND DATA PROCESSING}

FAST is located in Guizhou, China with its whole
aperture of 500\,m and an illuminated sub-aperture of 300\,m during normal operation.
A 19-beam receiver covering 1.05-1.45 GHz provides two data streams ({one for each linear polarization}; ~\citealt{Jiang19}). The observation was
carried out using the central beam of the 19-beam receiver on 25th Aug 2020. The data were captured by a digital backend based on Reconfigurable Open-architecture Computing Hardware–version2 (ROACH2) and recorded in search mode PSRFITS format with four polarizations, 8-bit, 49.152 \,$\mu$s sampling interval, and 4096 frequency channels, respectively. A total of 7139\,s of observations were recorded.

The data was processed to remove dispersion delay caused by the interstellar medium and folded modulo the predicted pulse period and integrated every 1\,s using {\sc dspsr}~\citep{Straten11}.
 {\sc psrchive} program {\sc paz} and {\sc pazi}~\citep{Hotan04} was used to flag and remove narrow-band and impulsive radio-frequency interference (RFI) of the data.
A polarization calibration noise signal was injected and recorded after the pulsar observation.
Polarization calibration was achieved by correcting for the differential gain and phase between the receptors through separate measurements using the noise diode signal. Flux density was calibrated using observations of 3C 286. Rotation measure (RM) was measured using the {\sc psrchive} program {\sc rmfit}~\citep{Hotan04}.
The dispersion measure (DM) for each sub-integration was measured using the {\sc tempo2} software package~\citep{Hobbs06}.
{More detail on obtaining DM is given in Section 3.3}.

\section{RESULTS}

\subsection{Emission variations near the eclipse}

\begin{figure*}
\centering
\includegraphics[width=170mm]{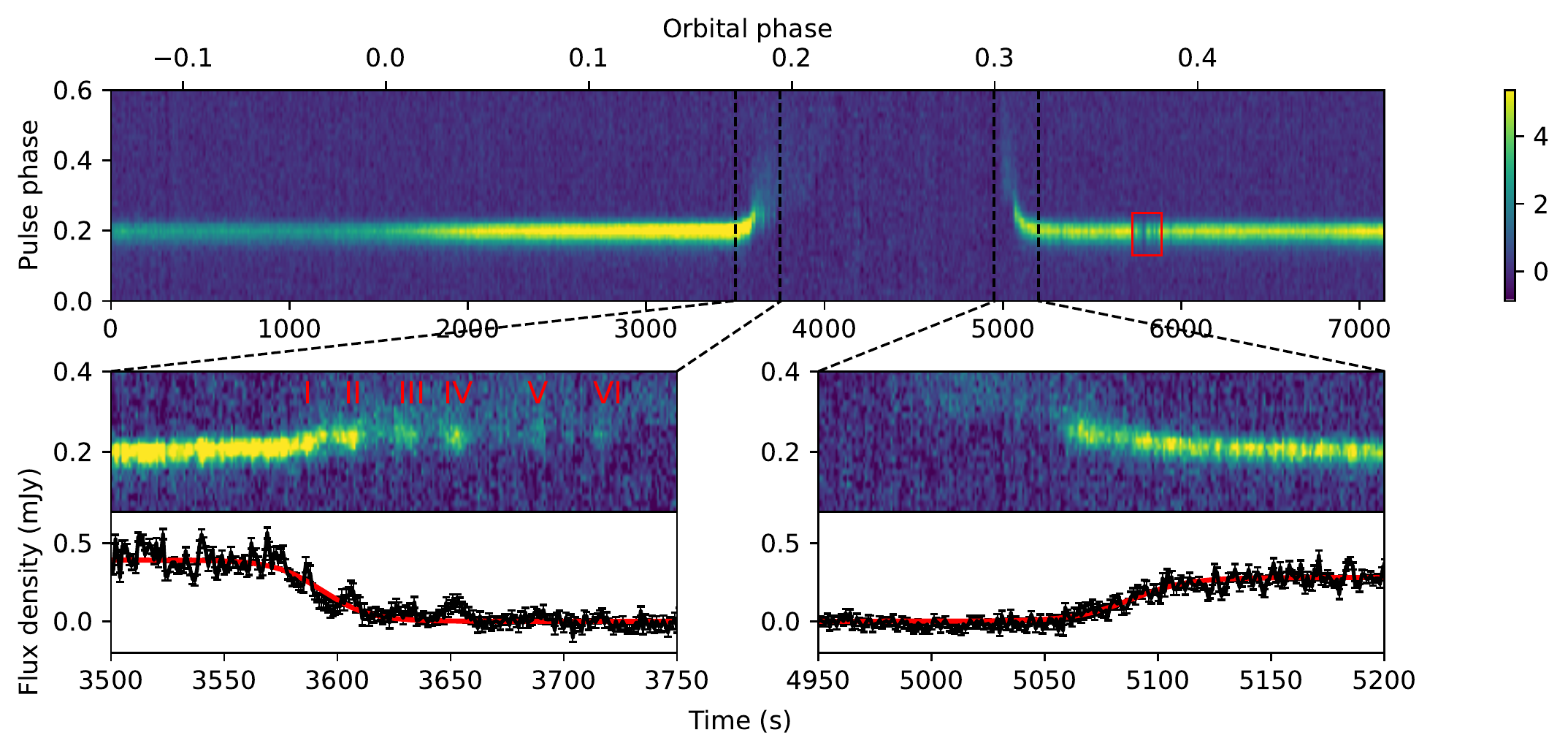}
\caption{The upper panel shows the total intensity of pulse emission vs. pulsar spin and orbital phases with the sub-integration of 1\,s of PSR J1720$-$0533. 
A mini-eclipse is labeled as the red box.
Enlargements of the ingress and egress of the pulsar are shown in the middle left and middle right panels, respectively. In the middle left panel, these six bright clusters are labeled as ``I'', ``II'', ``III'', ``IV'' , ``V'' and ``VI'', respectively. 
The bottom panels show the pulse flux density variations near the eclipse. The red curves show least-squares fit of Fermi–Dirac functions to the ingress and egress of the eclipse.}
\label{gray}
\end{figure*}

\begin{figure}
\centering
\includegraphics[width=80mm]{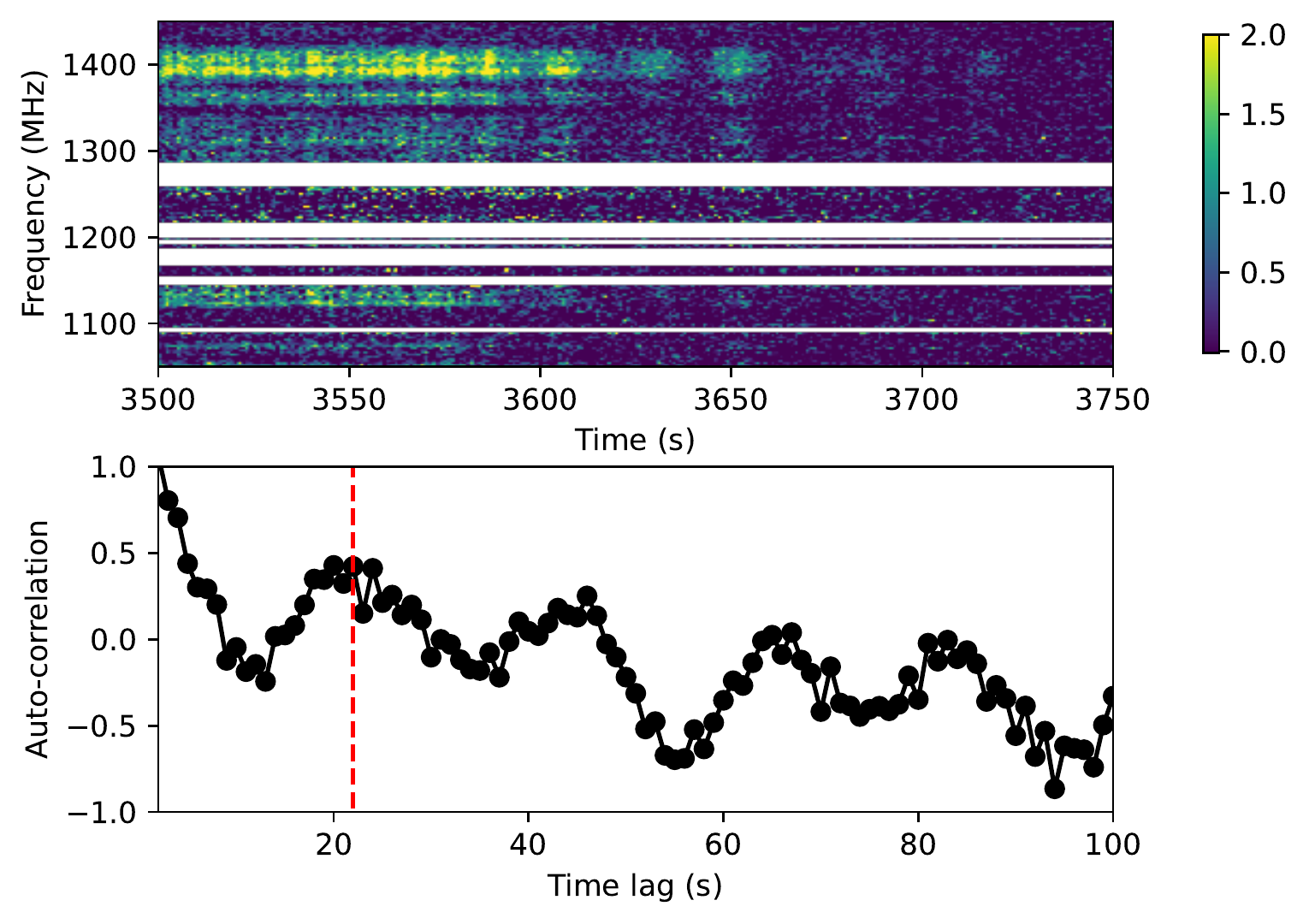}
\caption {Upper panel: the dynamic spectrum during the ingress of PSR J1720$-$0533.  Note that the frequency was binned into 256 channels and some channels are zapped because of RFI. Bottom panel: The auto-correlation function for the difference between the pulse flux density and the Fermi–Dirac fitting results during the ingress of the eclipse. The red dashed line is for the lag of 22\,s with the Pearson correlation coefficient of about 0.42.}\label{dy}
\end{figure}

The averaged pulse emissions of PSR J1720$-$0533  with sub-integration of 1\,s are shown in the upper panel of Figure \ref{gray}. {The pulse intensity becomes weaker, variable dispersion is seen through the shift of the pulse profile, and the profile is visibly scattered at the eclipse boundary.}
The pulsar underwent a strong scintillation that began around { orbital phase} of 0, which strengthened the emissions during the eclipse. {Note that this is interstellar scintillation, which is unrelated to the eclipse or eclipsing material.}
To measure the eclipse duration, Fermi-Dirac function was used to fit the {pulse flux density} variations during the ingress and egress. {More detail on the fitting is shown in Section 3.2. 
The flux density was measured using the {\sc psrchive} software package.
The eclipse duration was taken as the width of the eclipse at half maximum of the flux density.} 
The eclipse lasts about {1495$\pm$2\,s} which accounts for about {13.14\%} of the orbital period.
A mini-eclipse is seen at the orbit phase of 0.37 (the red square in the upper panel of Figure \ref{gray}) with duration about dozens of seconds. The mini-eclipses generally occur at different orbit phases in spider pulsars, which were attributed to clumps of plasma surrounding the eclipse medium~\citep{Deneva16, Polzin19}.

Enlargements of the ingress and egress of the eclipse are shown in the middle panels in Figure \ref{gray}. 
We found that the pulse emission of the pulsar during the ingress shows significant modulations and there are {six} bright emission clusters which are labeled as ``I'', ``II'', ``III'', ``IV'', ``V'' and ``VI'', respectively (see the middle left panel in Figure \ref{gray}).
{No such periodical modulations} are detected during the egress of the eclipse (see the middle right panel in Figure \ref{gray}). 
 {The flux densities during the ingress and engress gradually decreased and increased, respectively. 
 The duration of the ingress lasts longer than the egress and therefore the eclipse of PSR J1720$-$0533 is asymmetric.} The asymmetry eclipse of spider pulsars may be due to cometary-like tail of the intrabinary material that results from the orbital motion of the companion~\citep{Fruchter88,Main18}. The tail generally leads to eclipse egress lasting substantially longer than ingress~\citep{Tavani91}.

The dynamic spectrum during the ingress of PSR J1720$-$0533 are shown in the upper panel of Figure~\ref{dy}.
{These six bright emission clusters are also clearly seen in the dynamic spectrum and they are unrelated to the interstellar scintillation.}
Similar phenomenon has also seen in a BW PSR J2051$-$0827, {which maybe result from plasma lensing by the companion's material~\citep{Lin21}}. 
The plasma lensing possibly occurs during the eclipse ingress of PSR J1720$-$0533 as well.
We then investigated the periodicity of the modulations during the ingress of the eclipse. 
We subtracted the Fermi–Dirac fitting results from the flux density during the ingress and then calculated the auto-correlation function (the bottom panel of figure~\ref{dy}). We found that the correlation coefficient {reaches} its maximum at the time lag of 22\,s, which suggests that the modulations during the ingress of eclipse are quasi-periodic with {a period of $\sim$ 22\,s.}
{The { Fermi-Dirac} fit are used to estimate the average magnification of lensing and we obtained that the magnification is in the range of about 0.4$-$1.6 during the eclipse ingress.}

\subsection{Frequency-dependent eclipse}

\begin{figure}
\centering
\includegraphics[width=80mm]{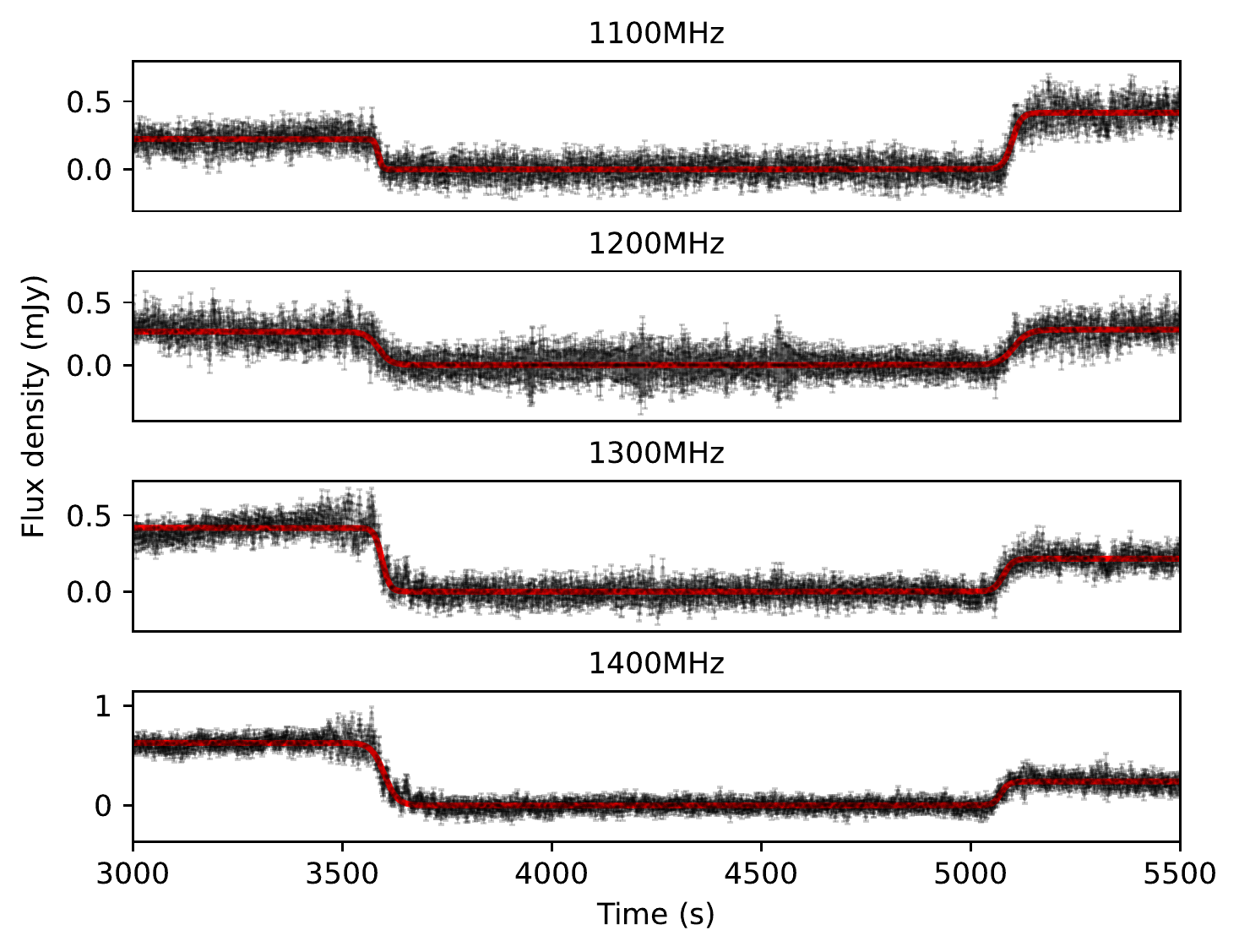}
\caption{Flux density variations for the eclipse of PSR J1720$-$0533 at different frequencies. The red line shows the least-squares fit of Fermi-Dirac function to the ingress and egress at each frequency sub-band.}\label{dy}
\end{figure}

To investigate the frequency dependence of the eclipse, we divided the entire band into four equal sub-bands with the central frequencies of 1100\,MHz, 1200\,MHz, 1300\,MHz and 1400\,MHz, respectively. {
The duration of eclipse was taken as the full width at half-maximum of the flux density and was calculated by fitting the ingress and egress flux densities with Fermi–Dirac functions $f=A{(e^{\frac{\phi-p_{1}}{p_{2}}}+1)}^{-1}$ with the amplitude A, the time at half-max of the pulse flux density $P_1$ and the slope $P_2$, was used to fit the flux density variations for the eclipse at each sub-band \bf \citet{Polzin18} . 
}
{The eclipse durations at these four sub-bands are 1513$\pm$3\,s, 1517$\pm$7\,s, 1483$\pm$3\,s and 1472$\pm$2\,s, respectively. }
Like other spider pulsars, the eclipse of PSR J1720$-$0533 lasts longer at lower frequency. However, the eclipse duration is not {monotonically} decreasing with increasing frequency because of the effect of scintillation which makes it is hard to confirm where the eclipse begins or ends. Our results are limited by the observational frequency bandwidth and observations with  ultra-wideband receiver will provide more information on the eclipse. 
We used a power-law function to fit the eclipse duration variations with increasing frequency and obtained that {the index $\alpha=-0.14\pm0.07$}, which is flatter than that of PSR B1957+20 of $-0.41\pm0.09$~\citep{Fruchter88} or PSR J1810+1744 of $-0.41\pm0.03$~\citep{Polzin18}.
Power-law functions was also used to fit both ingress and egress durations, and we obtained that {the indices for ingress and egress are  $-0.07\pm0.02$ and $-0.18\pm0.07$, respectively. The duration of egress are more frequency-dependent than egress.}

\subsection{DM and polarization profile variations}

\begin{figure}
\centering
\includegraphics[width=80mm]{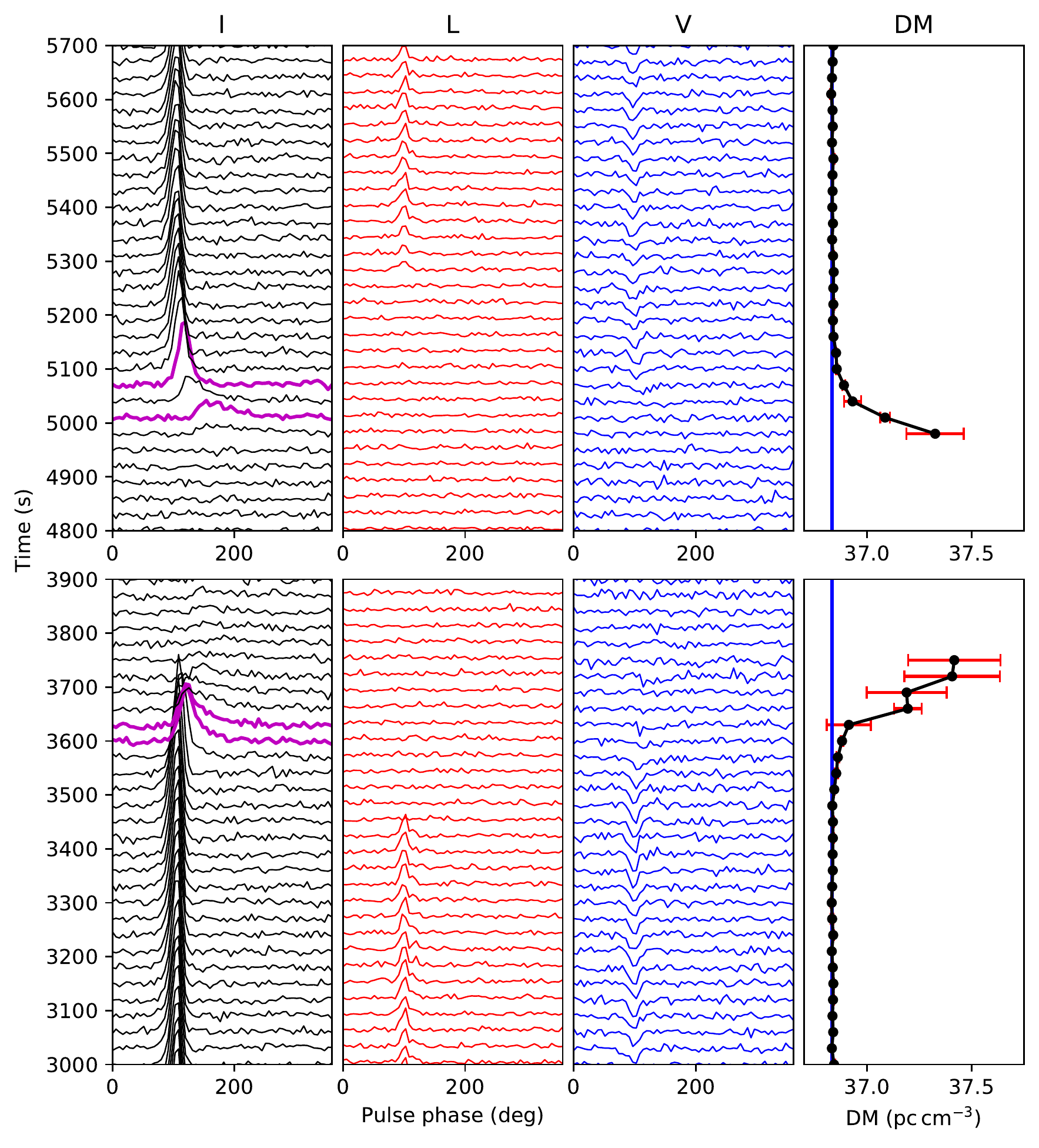}
\caption{Total intensity, I (black), linear polarization, L (red), and circular polarization, V (blue), average pulse profiles of PSR J1720$-$0533 during the ingress (bottom panels) and egress (upper panels) of the eclipse. The magenta lines are the the last and first pulses before and after the eclipse. The corresponding DM (black dots) variations of the eclipse are shown in the right panels. The red bars are the DM uncertainties. The horizontal blue line is for the DM of $36.8337\pm0.0006\,\rm cm^{-3}\,pc$ which is obtained by fitting the average out-of-eclipse profile.}
\label{rmdm}
\end{figure}

\begin{figure}
\centering
\includegraphics[width=80mm]{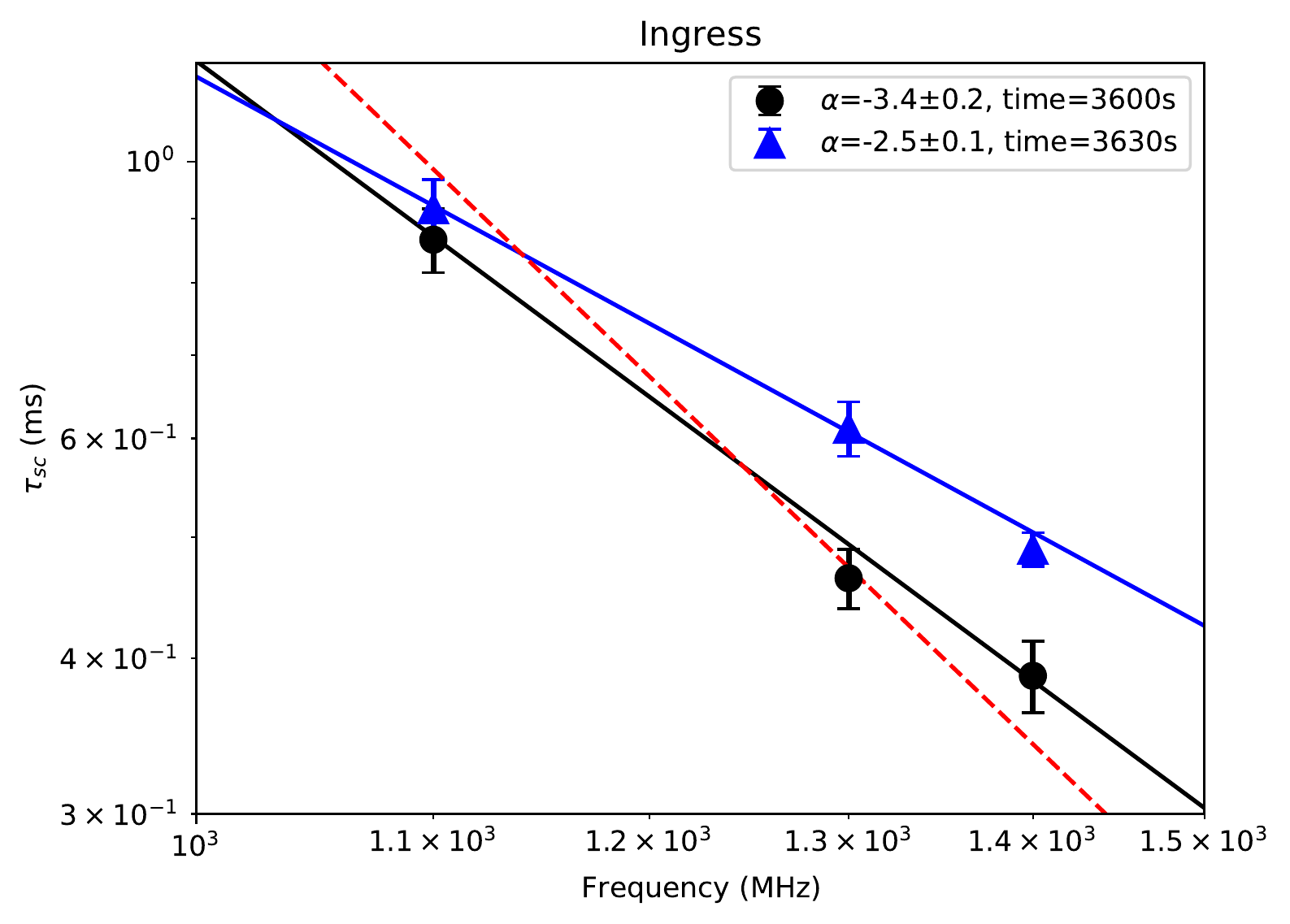}
\includegraphics[width=80mm]{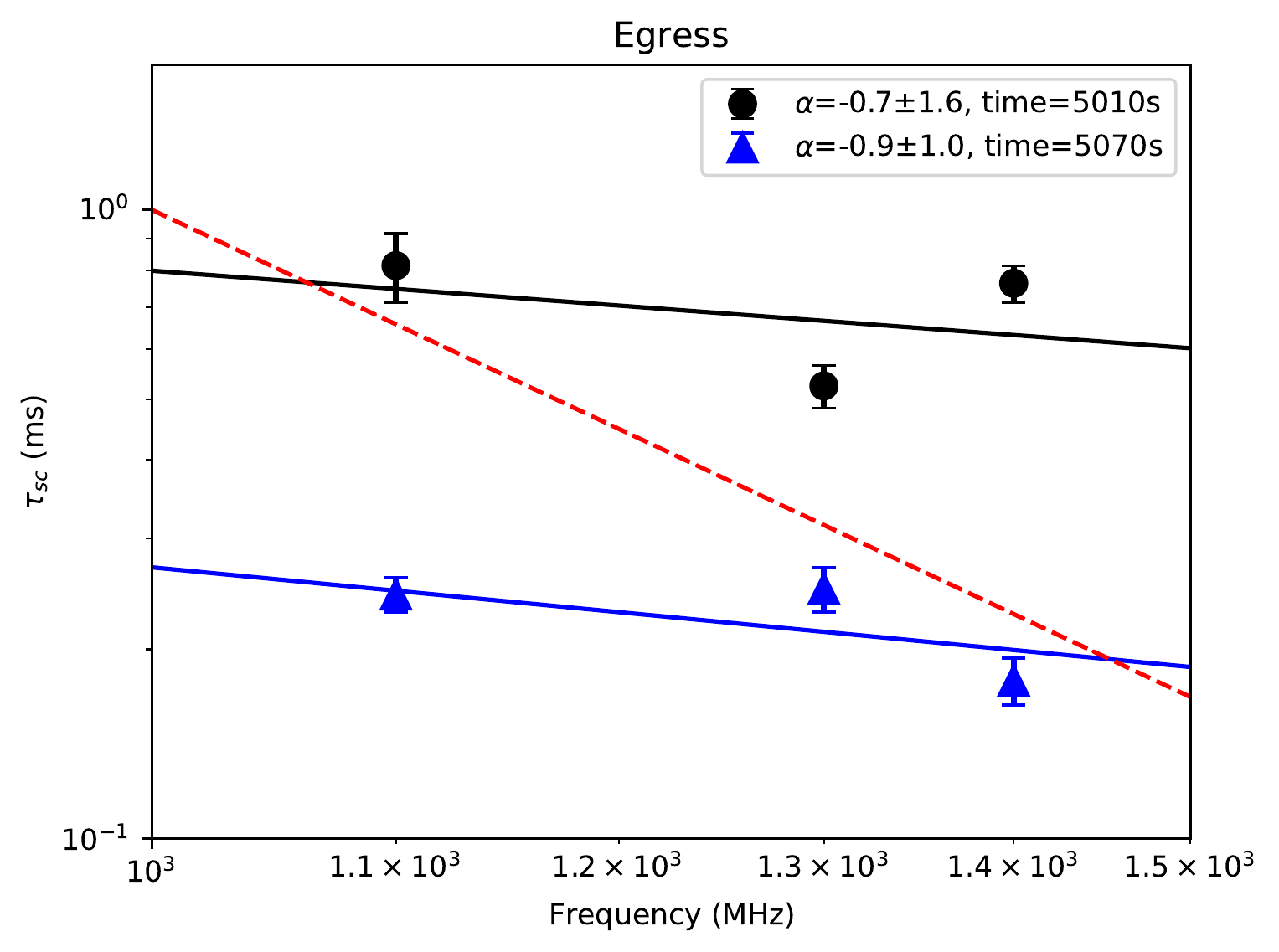}
\caption{The scatter broadening times $\tau_{sc}$ at different frequencies for PSR J1720$-$0533. The solid lines are the power law fitting with the index $\alpha$ and the dashed lines shows the thin-screen Kolmogorov prediction with the index $\alpha$=-4.4~\citep{Lee76}. Note that we did not fit the profiles at 1200\,MHz because of the limited S/N.}\label{scatter}
\end{figure}

{{\sc TEMPO2} software package~\citep{Hobbs06} were used to obtain the DM during the eclipse.} The {data} is folded with sub-integration of 30\,s. 
Then, we scrunched our data to 4 frequency channels with bandwidth of 100\,MHz.
A noise-free template was formed by fitting the integrated profile of the entire out-of-eclipse observation and ToAs were formed by cross-correlating pulse profiles with the standard template.
We fitted the DM of each sub-integration during the ingress and egress of of the pulsar, and the DM variations are shown {in the right panels of }Figure~\ref{rmdm}.
We also fitted the DM of the average out-of-eclipse profile and obtained a DM of $36.8337\pm0.0006\,\rm cm^{-3}\,pc$. 
As expected, the DM near the eclipse increases significantly and the maximum DM variations ($\Delta$DM) during the ingress and egress of the eclipse are about $0.6 \pm 0.2\,\rm cm^{-3}\,pc$ and $0.5 \pm 0.1\,\rm cm^{-3}\,pc$, respectively.

By analyzing the data with sub-integration of 30\,s, the polarization profiles of the pulses near the eclipse of PSR J1720$-$0533 are shown in Figure~\ref{rmdm}. 
{Note that the out-of-eclipse RM of $21\pm1\,\rm rad\,m^{-2}$ was used to correct the profiles.} 
We found that the linear polarization disappears earlier than both the total and circular polarized profiles during the ingress and it appears later during the egress. 
The times when the linear polarization disappears and appears are 3450\,s and 5250\,s, respectively, but the profiles of the total intensities at these times do not show significant variations. 
The corresponding $\Delta$DMs are $0.003\pm0.004\, \rm cm^{-3}\,pc$ and $0.005\pm0.001\, \rm cm^{-3}\,pc$, respectively, {which does not show significant variations.}

{The pulse profiles near the eclipse become wider than that of out-of-eclipse (the left panels of Figure~\ref{rmdm}).  
{ The dispersion smearing across each frequency channel is about 19\,$\mu$s for PSR J1720$-$0533, which can be negligible compared to the pulse profile.}
The scattering of the eclipse medium gives rise to an exponential decay of the pulse ~\citep{Williamson72}. 
{ We used the measured DM value to correct the profile, and then} divided the entire band into four equal sub-bands with the central frequencies of 1100\,MHz, 1200\,MHz, 1300\,MHz and 1400\,MHz, respectively and used a convolution of a Gaussian function with the exponential decay to fit the profile at each bands.
We choose four bright profiles near the eclipse at the times of 3600\,s, 3630\,s 5010\,s, and 5070\,s, respectively (the magenta lines in Figure~\ref{rmdm}). 
The $\tau_{sc}$ variations with frequency of these four profiles are shown in the upper and bottom panels of Figure~\ref{scatter}, respectively; and the power law fitting are shown by solid lines and the dashed lines shows the thin-screen Kolmogorov prediction with the index $\alpha$=-4.4~\citep{Lee76}. 
Note that we did not fit the profiles at 1200\,MHz because of the limited signal to noise ratio (S/N).
As seen, the scatter broadening time becomes larger while the index becomes smaller during ingress. However, the index dose not show significant variations during egress. 
{ 
\citet{Polzin20} showed evidence for the presence of an eclipse mechanism which only smears out pulsations while does not remove flux for PSR J1816+4510, and attributed this to scattering in outflowing material from the companion. It is also a possible scenario for PSR J1720$-$0533.} 
The scattering maybe also play an important role in the eclipse mechanism of PSR J1720$-$0533.
}

\section{ DISCUSSION AND CONCLUSIONS}

We present an high sensitive observation of a newly discovered black widow PSR J1720$-$0533 using FAST. The scintillation maximum throughout the eclipse in our observation provides an opportunity to study the emission variations near the eclipse in detail.
We found there are quasi-periodic pulse intensity modulations with { a period of $\sim$22\,s} during the ingress of the eclipse. 
No such emission modulations are detected during the egress of the eclipse. 
The eclipse is asymmetric and the duration of ingress is longer than egress.
The modulation during the ingress shows similar properties to the highly variable emissions throughout the eclipse of PSR J2051$-$0827, which are attributed to lensing by the intrabinary material~\citep{Lin21}.
The phenomenon of PSR J2051$-$0827 demonstrates a link between DM and lensing. 
We suggested the plasma lensing possibly occurs in PSR J1720$-$0533 as well. 
Unfortunately, the DM variations of each modulation cluster during the  ingress of PSR J1720$-$0533 cannot be measured precisely because of the limited S/N. 

For PSR J2051$-$0827, the radiation beam sweeps across the edge of the eclipsed medium~\citep{Stappers01} and the emissions do not completely disappear during the eclipse. 
The plasma lensing occurs throughout the entire eclipse and the maximum $\Delta$DM during the eclipse is about 0.07\,$\rm cm^{-3}\,pc$~\citep{Lin21}. 
However, for PSR J1720$-$0533, the emission is blocked by the eclipsed medium completely during the eclipse with much longer eclipse duration and the plasma lensing only occurs during the ingress of the eclipse. 
The maximum $\Delta$DM during the eclipse ingress is about $0.6\pm0.2\,\rm cm^{-3}\,pc$ which is similar to that during the egress of $0.5\pm0.1\,\rm cm^{-3}\,pc$. 
{If the DM variations are tied to the flux density variations~\citep{Lin21}, the different DM variations during egress may account for the different timescale compared to ingress. }

Besides PSR J1720$-$0533 and PSR J2051$-$0827, plasma lensing has been detected in two other spider pulsars: PSR B1744$-$24A~\citep{Bilous19} and PSR B1957+20~\citep{Main18} with different manifestations.
Unusually bright single pulses are detected largely  near eclipse ingress and egress of the two pulsars, have intensities up to dozens of times than the average pulse intensity, and pulse shape similar to that of the average pulse profile.
It is difficult to explain these bright pulses via
scintillation in the interstellar medium, as a separate emission mode, or as conventional giant pulses.
The authors suggested these bright pulses are attributed to the lensing by the eclipsed medium~\citep{Main18, Bilous19}. 
The duration of theses bright pulses is about {dozens} of millisecond which is much shorter than that seen in PSR J1720$-$0533 and PSR J2051$-$0827 with duration of dozens of seconds. 

{To estimate the size and location of plasma lens, a single 1D lens model of \citet{Cordes17} was used and the DM within the lens was assumed to follow Gaussian distribution.
The size of the lens is $a_{\rm lens}=a R_{\rm sep}$ and the distance from the pulsar to the lens is $d_{\rm sl}=d R_{\rm sep}$ with the separation between pulsar and companion $R_{\rm sep}$ and the dimensionless quantities a and d. 
The Fresnel scale at the lens plane is $r_{\rm F}\approx \sqrt{c d_{\rm sl} d_{\rm lo}/\nu d_{\rm so} }$, where $\nu$ is the observation frequency, $c$ is the velocity of light, $d_{\rm lo}$ and $d_{\rm so}$ are the distance from the observer to the lens and from the observer to the pulsar, respectively. 
We assume $d_{\rm so} \approx d_{\rm lo}$ because the lens is much closer to the pulsar than to the observer. 
The maximum pulse amplification is $G\sim a_{\rm lens}/r_{\rm F}$. The time of caustic crossing $t_{\rm c}\sim a_{\rm lens}(\delta G/G)/v_{\rm trans} G^2 \cdot d_{\rm lo}/ d_{\rm so}$ with the effective transverse velocity $v_{\rm trans}$. 
We take $v_{\rm trans}$ as the orbital velocity of the companion, $G=1.6$, $\delta G/G \approx 1$ and $t_{\rm c} \approx 20$\,s,  and obtained that $a\approx 2.6\times10^{-2}$ and $d\approx 9.8\times10^5$. The corresponding lens size $a_{lens}\approx 2.3\times10^4$\,km.

}

The pulse emission of PSR J1720$-$0533 becomes depolarized near the eclipse, like PSR J1748$-$2446A~\citep{You18} and PSR J2256$-$1024~\citep{Crowter20}.
The depolarization occurs when the RM shows rapid time variations~\citep{You18}. 
The fluctuations of the DM and/or parallel component of the magnetic field in the eclipse medium can result in rapid RM variations.
For PSR J1720$-$0533, the DM variations at the orbital phases where the linear polarization disappears during the eclipse ingress and egress are $0.003\pm0.004\, \rm cm^{-3}\,pc$ and $0.005\pm0.001\, \rm cm^{-3}\,pc$, respectively, {which is almost unchanged.} 
Therefore, the RM variations  are  more likely results from the fluctuations of the parallel component of the magnetic field in the eclipse medium and causes the depolarization of PSR J1720$-$0533. 
Our results suggest that there may be a significant magnetic field in the eclipse medium.

We then estimated the magnetic field strength of the eclipse medium. 
{For spider pulsars, the pressure of the pulsar wind $P_{\rm pw}=\pi I \dot{P}/c a^2 P^3$ with the pulsar  period $P$, the derivative of period $\dot{P}$, the pulsar’s inertia moment $I$, the orbital separation $a$ and the speed of light $c$. The magnetic pressure of the companion wind $P_{\rm cw}=B_{\rm E}^2/8 \pi$ with the magnetic field strength $B_{\rm E}$. 
If the pulsar wind pressure equals the magnetic pressure of the companion wind at the interface, $P_{\rm pw}=P_{\rm cw}$~\citep{Thompson94}, $B_{\rm E}=51P_{-3}^{-3/2}\dot{P}_{-20}^{1/2}a^{-1}_{11}$ taking $I=10^{45}\,{\rm g cm^2}$, where $a_{11}$ is the orbital separation in the unit of $10^{11}$\,cm, $P_{-3}$ and $\dot{P}_{-20}$ are the spin period in the unit of $10^{-3}$\,s and the time derivative of period in the unit of $10^{-20}$\,s/s. For PSR J1720$-$0533, $P$=3.26\,ms, $\dot{P}=7.46\times10^{-21}$\,s/s and $a$=1.3\,${\rm R_{\odot}}$, the implied magnetic filed of the eclipse medium $B_{\rm E}\approx 8$\,G.

}

We also studied the effect of scattering near the eclipse of PSR J1720$-$0533. We found that the scatter broadening time is frequency-dependent during the ingress. 
Our results suggest that the scattering plays an important role in the eclipse at 1250\,MHz for PSR J1720$-$0533.
Our result is consistent with the conclusion of \citet{Thompson94} that pulse smearing is more important for the eclipse at higher frequency for spider pulsars. 
{ \citet{Polzin20} presented observations with the pulsed and imaged continuum fluxes simultaneously and found that the continuum flux of PSR J1816+4510 is consistent with the pulsed flux during ingresses, but it extends to a significantly early orbital phase than the corresponding pulses flux in one egress of the eclipse.
They suggested that this maybe result from scattering in outflowing  material from the companion, and it seems that the scattering only occurs during the eclipse egress of PSR J1816+4510.    
For PSR J1720$-$0533, the scattering maybe only occurs during the ingress in our observation and further observations with the pulsed and continuum fluxes simultaneously using FAST will provide more information on the eclipse.}

\section*{Acknowledgments}

This is work is supported by the the National SKA Program of China (No.2020SKA0120100), National Natural Science Foundation of China (No.12041304, 12041303, 12163001), the National Key Research and Development Program of China (2017YFA0402600), the Youth Innovation Promotion Association of Chinese Academy of Sciences, the 201* Project of Xinjiang Uygur Autonomous Region of China for Flexibly Fetching in Upscale Talents, the Operation, Maintenance and Upgrading Fund for Astronomical Telescopes and Facility Instruments, budgeted from the Ministry of Finance of China (MOF) and administrated by the Chinese Academy of Science (CAS), and the Key Lab of FAST, National Astronomical Observatories, Chinese Academy of Sciences. This work made use of the data from the Five-hundred-meter Aperture Spherical radio Telescope, which is a Chinese national mega-science facility, operated by National Astronomical Observatories, Chinese Academy of Sciences. The authors wish to thank K. Liu and  F.X. Lin for useful suggestions and comments that improved the manuscript.

\software{DSPSR \citep{Straten11}, PSRCHIVE \citep{Hotan04} and TEMPO2 \citep{Hobbs06}}

\bibliography{sample63}{}
\bibliographystyle{aasjournal}

\end{document}